# Universal computation with limited resources: Belousov-Zhabotinsky and Physarum computers

Andrew Adamatzky[1], Ben De Lacy Costello[2] and Tomohiro Shirakawa[3]

1. Faculty of Computing, Engineering and Mathematical Sciences, University of the West of England, Bristol, United Kingdom
2. Faculty of Applied Sciences, University of the West of England, Bristol, United Kingdom
3. Department of Computational Intelligence and Systems Science, Tokyo Institute of Technology, Tokyo, Japan

## Abstract

Using the examples of an excitable chemical system (Belousov-Zhabotinsky medium) and plasmodium of *Physarum polycephalum* we show that universal computation in a geometrically unconstrained medium is only possible when resources (excitability or concentration of nutrients) are limited. In situations of limited resources the systems studied develop travelling localizations. The localizations are elementary units of dynamical logical circuits in collision-based computing architectures.

## 1. Introduction: architecture-based vs architecture-less computing

The first thoughts on the implementation of computational operations with patterns propagated in spatially extended non-linear systems date back to 1800's where Plateau experimented with the problem involving the calculation of the surface of smallest area bounded by a given closed contour in space [Courant and Robins, 1941] (the classical problem of calculating a minimal spanning tree of planar points using a soap film). These ideas were rediscovered many times but mostly in a framework of theoretical design of novel algorithms, of these the grass fire transformation [Blum, 1968; Calabi & Hartnett, 1968] is the most famous example. In 1990's the new field of reaction-diffusion computing — computation with excitation and diffusive waves in two-dimensional chemical media — was conceived in experiments using the Belousov-Zhabotinsky medium [Kuhnert & Agladze, 1989; Rambidi, 1998] and precipitating chemical systems [Tolmachiev & Adamatzky, 1996].

A reaction-diffusion computer is a spatially extended chemical system, which process information using interacting growing patterns, excitable and diffusive waves [Adamatzky, 2001; Adamatzky, De Lacy Costello & Asai, 2005]. In reaction-diffusion processors, both the data and the results of the computation are encoded as concentration profiles of the reagents. The computation is performed via the spreading and interaction of wave fronts. Reaction-diffusion chemical processors are now 'classical' examples of non-linear medium computers.

A non-linear medium processor can be either specialised or general-purpose (universal). A specialized processor is built to solve only one particular problem, possibly with different data sets and with variations in the interpretation of the results. Specialised computing devices are quite handy when we deal with image processing, problems of mathematical morphology, or computation on graphs [Adamatzky, 2001; Adamatzky, De Lacy Costello, Asai, 2005]. A device is called computationally universal if it computes a functionally complete set of logical operations. To prove a medium's universality one must represent quanta of information, routes of information transmission and logical gates, where information quanta are processes, in states of the given system. This can be done in two ways: architecture-based and architecture-less.

An architecture-based, or stationary, computation implies that a logical circuit is embedded into the system in such a manner that all elements of the circuit are represented by the system's stationary states. The architecture is static. If there is any kind of 'artificial' or 'natural' compartmentalisation the medium is classified as an architecture-based computing device. Personal computers, living neural networks, cells, and networks of chemical reactors are typical examples of architecture-based computers. Until very recently most experimental realizations of chemical computers were based on geometrically constrained media:

waves propagate in channels and interact at the junctions between the channels. The approach is practical but also acts as a cul-de-sac of novel computing, because it is simply an implementation of conventional computer architectures in new materials.

The ideas of colliding signals were established in an automata framework in around 1965, in cellular automaton algorithms for multiplication [Atrubin, 1965] and generation of prime numbers [Fisher, 1965]. In 1982 Berlekamp [1982], Conway and Gay [1982] proved that the Game of Life cellular automaton can imitate a Turing machine. Nearly at the same time Fredkin and Toffoli [1978] showed how to design a non-dissipative computer that conserves the physical quantities of logical signal encoding and information in the physical medium. They further developed these ideas in the conservative logic [Fredkin & Toffoli 1982], a new type of logic with reversible gates. Thus, a concept of computing that explores elastic collisions with balls and fixed reflectors — the billiard ball model (BBM) was established in which they demonstrated that given a container with balls it is possible to implement any kind of computation, i.e. they demonstrated the logical universality of the BBM. The BBM has been ingeniously implemented in 2D cellular automata by Margolus [1984].

A collision-based, or dynamical, computation employs mobile compact finite patterns, mobile self-localized excitations [Adamatzky, 2004; De Lacy Costello & Adamatzky, 2005] or simply-localizations, in active non-linear medium. The localizations travel in space and computation occurs when they collide with each other. The essential features of collision-based computing are as follows. Information values (e.g. truth values of logical variables) are given by either absence or presence of the localisations or other parameters of the localisations. The localisations travel in space and computation occurs when they collide with each other. There are no predetermined stationary wires (a trajectory of the travelling pattern is a transient wire). Almost any part of the medium space can be used as a wire. Localisations can collide anywhere within a space sample, there are no fixed positions at which specific operations occur, nor location specified gates with fixed operations. The localisations undergo transformations (e.g. change velocities, form bound states, annihilate or fuse when they interact with other mobile patterns. Information values of localisations are transformed as a result of collision and thus a computation is implemented [Adamatzky, 2003].

Localizations in excitable chemical media are very sensitive objects: they stay localized for a very short time and then either expand or collapse, they are highly sensitive to the physical conditions of the medium, and, in principle, not easy to generate and purposefully direct their motion [Toth et al., 2007]. For this reason we have started to look for a feasible alternative to chemical localizations, and tried to locate a non-linear medium primitive enough to be described by basic rules of reaction and diffusion, exhibiting rich dynamics of localizations and stable enough to support the localisations for a substantial period of time. Encapsulating a reaction-diffusion chemical system in a membrane would certainly stabilize its behaviour.

Thus we have come to the conclusion that a vegetative state, or plasmodium, of *Physarum polycephalum* is to date the best representation of a non-linear medium encapsulated in an elastic membrane. In the present paper we compare the behaviour and functionality of localizations in the Belousov-Zhabotinsky reaction and also a plasmodium of *P. polycephalum* in order to evaluate the plasmodium's suitability for collision-based universal computing. We also specify the basic conditions of a non-linear medium which lead to the emergence of localizations.

The paper is structured as follows. Section 2 demonstrates how excitability affects morphology of wave-fronts in discrete excitable lattices. Emergence of localized wave-fragments in a numerical model of excitable chemical systems is discussed in Sect. 3. Experimental results on development of localizations in plasmodium of *Physarum polycephalum* are provided in Sect. 4. Comparative analysis of localizations from an unconventional computing point of view is proposed in Sect. 5.

## 2. Localizations on discrete excitable lattices

As have been already outlined in our previous work (see review in [Adamatzky, 2001]) localizations in excitable lattices emerge at the narrow zone of excitability, just near the threshold of excitability.

Consider two-dimensional excitable lattice, where every site takes three states: resting, excited and refractory. All sites update their states in parallel. Transitions from excited to refractory and from refractory to resting are unconditional, they take place independently on the neighbourhood configurations. In such a model localizations emerge when every resting site becomes excited only if it has two neighbours in the excited state (see example snapshot of the medium in Fig. 1). A resting site becomes excited depending on the number of excited neighbours. All cells update their states simultaneously, in discrete time and depending on the states of their closest neighbours, e.g. the localization $\frac{-|-}{+|+}$ is travelling South. In this example excitability is limited 'from above', i.e. the number of excited neighbours capable of exciting a resting cell is capped by two. If we would allow a resting state to take excited state if the number of neighbours simply exceeded two, then 'classical' circular and target waves would emerge in the excitable lattice.

A growth similar to an amoeboid can be imitated in a three-state, eight-cell neighbourhood, two-dimensional cellular automaton where an excited cell does not necessarily take the refractory state, the so-called model of retained excitation [Adamatzky, 2007]. In this cellular automaton every cell takes resting, excited and refractory states. All cells update their states simultaneously, in discrete time and depending on the states of their closest neighbours.

A resting cell makes the transition to an excited state if the number of excited neighbours belongs to the excitation interval $[\theta_1, \theta_2]$. An excited cell remains excited if the number of excited neighbours belongs to the interval $[\delta_1, \delta_2]$ of excitation retention. Each rule can be described as a tuple $R(\theta_1, \theta_2, \delta_1, \delta_2)$. In the local transition rules space we have located [Adamatzky, 2007] the domain $\mathbf{R}(3,3,\delta_1,\delta_2)$, $\delta_1 \in [1,3]$ and $\delta_2 \in [5,8]$, where rules support growth of non-uniform patterns, with sometimes articulated directionality.

In cellular automata governed by functions from $\mathbf{R}(3,3,\delta_1,\delta_2)$. As demonstrated in Fig. 2, an initially circular domain of stimulation is transformed to an asymmetric and irregularly growing pattern. The growth rate of different parts of such configurations varies significantly, between 0.2 and 1.0. This causes a morphological heterogeneity. Wave-fronts are subdivided onto pseudopodia-like growing domains. In many cases directed, or almost directed, growth of pseudopodia is guided by small tips. In some cases the growing tips consist of excited and refractory states (Fig. 3a), and in other cases the tip is comprised of excited states only (Fig. 3b-c). Increase of the upper boundary $\delta_2$ of the excitation retention interval reduces branching, while increase of the lower boundary, $\delta_1$, compresses growing patterns. In cases of a very wide interval of excitation retention, $\delta_2 = 8$, the role of refractory states diminishes [Adamatzky, 2007a].

We demonstrated that in a cellular automaton model of a system with very limited excitability of elements coupled with their restricted abilities to stay excited, amoeboid patterns develop, with articulated growth-points and pseudopodia-like excesses.

In both previously discussed cases we demonstrated that *a mobile localization emerges in a discrete excitable lattice only if the excitability of the lattice sites is at the threshold level and limited from above* (i.e. too much excitation in a resting site's neighbourhood prevents the resting site from becoming excited).

This finding perfectly correlates with our recent computational experiments on evolving reaction-diffusion cellular automata supporting localizations [Adamatzky et al., 2007]. We studied hexagonal cellular automata with immediate cell neighbourhood and three cell-states {$A$, $B$, $S$}, where states $A$ and $B$ represent quasi-chemical reactants and $S$ is a quiescent state, or a substrate. Every cell calculates its next state depending on the integral representation of states in its neighbourhood, i.e. how many neighbours are in each of the states. Such a transition rule is somewhat analogous to a quasi-chemical reaction: $\alpha A + \beta B \rightarrow aA + bB + cS$ where $\alpha + \beta \leq 7$ and $a + b + c = 1$. We have employed evolutionary algorithms [Sapin et al., 2007] to breed local transition functions that support mobile localizations, and characterize sets of the functions selected (Fig. 4).

Analysis of the set of functions evolved allowed us to speculate that *mobile localizations are likely to emerge in the quasi-chemical systems with limited diffusion of one reagent, where a small number of* *molecules are required for amplification of travelling localizations, and reactions leading to stationary localizations involve relatively equal amounts of quasi-chemical species* [Adamatzky et al., 2007].

We do not offer examples of logical gates realizable with colliding discrete localizations because there is an abundance of examples in [Adamatzky, 2001].

## 3. Travelling localizations in chemical media: Belousov-Zhabotinsky reaction

Typically in a spatially extended Belousov-Zhabotinsky (BZ) medium local perturbations of the medium initiate a circular travelling wave if they exceed a reaction dependent threshold. If the perturbation is long lived then after a given refractory period for the reaction a second circular wave will be initiated leading eventually to the formation of classical target waves. In addition local perturbations can cause circular waves to break forming spiral waves to be generated. The circular waves travel outwards from the point of stimulation due to diffusion (Fig. 5a). Quite recently it was experimentally demonstrated that by using a light sensitive catalyst in the BZ reaction this allows for fine control of the excitability of the medium by using light [Beato & Engel, 2003; Brandstädter et al., 2000]. Light increases production of bromide ions, which are inhibitors of the reaction. The increase in the inhibitor concentration naturally leads to a decrease of the medium's excitability. The BZ reaction at the critical threshold of excitability exhibits travelling localizations, wave-fragments (Fig. 5b).

To inspect how decreases in excitability leads to the formation of travelling localizations in the BZ reaction we simulated the medium using a two-variable Oregonator model modified to account for photochemistry [Field & Noyes, 1973; Krug et al., 1990; Adamatzky, 2004]:

$$\frac{\partial u}{\partial t} = \frac{1}{\varepsilon}\left(u - u^2 - (fv + \Phi)\frac{u - q}{u + q}\right) + D_u \nabla^2 u$$

$$\frac{\partial v}{\partial t} = u - v.$$

The variables $u$ and $v$ represent the instantaneous local (dimensionless) concentrations of the bromous acid autocatalyst and the oxidized form of the catalyst, $HBrO_2$ and tris (bipyridyl) Ru (III), respectively. The ratio of the time scales of the two variables, $u$ and $v$, is represented by $\varepsilon$, which depends on the rate constants and reagent concentration; $f$ is a stoichiometric coefficient. The scaling parameter, $q$, depends on reaction rates. We assumed that the catalyst is immobilized in a thin-layer of gel therefore there is no diffusion term for $v$.

Parameter $\Phi$ is a rate of bromide production, induced by light, the parameter also represents the excitability of the system. A moderate intensity of light will facilitate the excitation process whereas a higher light intensity will produce excessive quantities of bromide which suppresses the reaction. *The BZ system was sub-excitable at $\Phi$=0.04. In the sub-excitable mode the model of the BZ system exhibits localized wave-fragments, which keep their shape for some period of time, and travel as compact patterns (similarly to dissipative solitons) until they collide with other localizations, impurities or just spontaneously lose their stability.*

By colliding these travelling localizations we can implement almost any type of logical operations in BZ medium [Adamatzky, 2004; De Lacy Costello & Adamatzky, 2005; Adamatzky and De Lacy Costello, 2007; Toth et al., 2007]. Figure 6 shows the dynamics of the wave-fragments implementing a collision-based gate, with two inputs ($x$ and $y$) and four outputs: two outputs $x \wedge y$, one output $x \wedge \neg y$ and one output $\neg x \wedge y$. The full range of collisions between mobile excitations, discovered in our computational models, has been verified experimentally in [De Lacy Costello & Adamatzky, 2005; Toth et al., 2007], therefore we do not provide chemical laboratory illustrations in the present paper.

## 4. Emergence of localizations in plasmodium of *Physarum polycephalum*

The plasmodium of *Physarum polycephalum* is a unicellular and multinuclear giant amoeba that is visible to the naked eye. This organism, formed by aggregation and fusion of a myriad of cells, is so to speak a 2-dimensional mass of cytoplasm encapsulated in a plasma membrane. The very large size of the plasmodium allows the single cell to be highly amorphous. The plasmodium shows synchronous oscillation of cytoplasm throughout its cell body, and oscillatory patterns control the behaviours of the cell. As a result, all the parts of the cell behave cooperatively, allowing us to use the organism in various computational tasks such as maze-solving [Nakagaki, 2000, 2001], construction of logical gates [Tsuda et al, 2004], formation of Voronoi diagram [Shirakawa & Gunji, 2007a], and robot control [Tsuda et al., 2007].

The oscillatory cytoplasm can be regarded as a spatially extended nonlinear excitable media, as many of the models of the plasmodium have indicated (for example, see [Matsumoto et al., 1988; Nakagaki et al., 1999; Yamada et al., 2007]. Here we demonstrate that the sheet of the plasmodium shows self-localization under certain experimental conditions and is also potentially useful as a source for collision-based computation.

In all of experiments demonstrated in this section, the plasmodia cultured on wet paper towels were scraped from the towels and then weighed (About the method for culture, see [Camp, 1936]]). Then 1 or 8 pieces of the plasmodia with constant weight (20 mg) were placed on an agar gel substrate in Petri dishes, which either contain nutrient or repellent.

When the plasmodium is cultivated on a nutrient-rich substrate (agar gel containing crushed oat flakes) it exhibits uniform circular growth (Fig. 7 right) similar to the excitation waves in the excitable BZ medium. If the growth substrate lacks nutrients, e.g. the plasmodium is cultivated on a non-nutrient and repellent containing gel (Fig. 7 left), a wet filter paper or even glass surface localizations emerge and branching patterns become clearly visible. The pseudopodium (Fig. 8) propagates in a manner analogous to the formation of wave-fragments in sub-excitable BZ systems.

Furthermore, in the experiments with plasmodium we have discovered that interaction between 'colliding' pseudopodia depends on the condition of the substrate, especially on the richness of nutrients in substrate the pseudopodia is relatively rich with nutrients (Fig. 9), pseudopodia fuse, and in poor substrate the pseudopodium spread homogeneously and densely, so pseudopodia cannot turn and they always fuse when they collide. On the contrary, the pseudopodium of Physarum in unfavorable conditions is not so dense spatially, so it can easily turn. Thus pseudopodia in unfavorable conditions may sometimes fuse and sometimes avoid each other.

Pseudopodium of Physarum shows two types of response when it collides with another pseudopodium or another individual. When two pseudopodia collide, they fuse if there is no space in which to move in order to avoid each other. They avoid each other and make a turn if there is enough space available. Physarum in favorable condition spread homogeneously and densely, so pseudopodia cannot turn and they always fuse when they collide. On the contrary, the pseudopodium of Physarum in unfavorable conditions is not so dense spatially, so it can easily turn. Thus pseudopodia in unfavorable conditions may sometimes fuse and sometimes avoid each other.

## 5. (Dis)advantages of encapsulation while implementing computation

In our previous studies we have speculated, and have provided some argument that the plasmodium of *P. policephalum* is a good real-world prototype of a reaction-diffusion excitable chemical medium encapsulated in an elastic membrane [Adamatzky, 2007a]. Now we will compare the behaviour of and interaction between mobile localizations in discrete excitable lattices, excitable chemical media, and plasmodium of *P. polycephalum*. We will also build up parallels between the degree of excitability of a chemical media and their discrete models, and the concentration of nutrients in *P. polycephalum* cultures.

**Excitable mode.** When in an excitable mode a medium reacts to local stimulation by generating target and spiral waves (Fig. 5a). Similarly, in a nutrient rich mode Physarum develops circular propagating patterns, analogues of target waves (Fig. 9).

Waves in excitable BZ medium annihilate when they collide (because each wave has a refractory tail). Physarum growing patterns do also fuse entirely (Fig. 9) in nutrient-rich conditions. The fusion may be prevented in certain conditions (Fig. 10) and the structure formed by the wave fronts are very similar to Voronoi diagrams in precipitating chemical processors [De Lacy Costello et al., 2004].Morphology of wave-fronts in BZ system and plasmodium is similar.

**Subexcitable mode.** Both the BZ reaction and plasmodium of *P. Polycephalum* exhibit mobile localizations with similar morphology, see Fig. 5b and Fig. 8. The following types of collision between wave-fragments can be observed in the BZ medium: reflection, attraction, repulsion, sliding and shifting-sliding (Fig. 11); see [Adamatzky & De Lacy Costello, 2007] for verification of collisions in computational models and [Toth et al., 2007] for verification in chemical laboratory experiments. Different behaviour of wave-fragments in collisions is due to angle of fragments and velocity vectors. However, most collisions of wave-fragments in BZ involve some annihilation of part of original waves, followed by fusion or reflection (unless fragments trajectories are changed by just proximity to other fragment, which is sometimes observed).

Are all these types of collision observed in the interaction of pseudopodia of Physarum's plasmodium? As with BZ wave-fragments, results of collision of Physarum pseudopodia depend on the orientation of their velocity vectors and the growth tips positions at the moment before collision. Usually, if localizations are just 'brushing' each other then reflection occurs.

Reflection and repulsion of pseudopodia are observed experimentally on a nutrient-poor substrate (analog of sub-excitable mode of BZ medium) (Fig. 10 and Fig. 12). Interaction of localizations in nutrient-poor substrate is visibly distinct from that on a nutrient-rich substrate (Fig. 9).

We never observed clear attraction: in situation where pseudopodia fuse (nutrient-rich), they usually just continue along their original trajectories till they come into physical contact with another pseudopodia. Neither sliding not shifting-sliding were observed in experiments with plasmodium.

However, it should be noted that it was also difficult to reproduce behaviour seen in theoretical chemical models in actual chemical experiments. Therefore it is likely with living organisms that the same difficulties of controlling exactly the exact trajectories, velocity vectors etc. may limit the observation of more subtle behaviours. This is not to say that they do not exist.

**Computing with localizations.** Wave-fragments in the BZ medium and tips of pseudopodia in Physarum can represent signals in collision-based computing schemes [Adamatzky, 2003]. Due to the ability of the medium to return to its original resting state, the BZ medium allows for --- theoretically --- unlimited numbers of signals to share the same loci of medium, if the signals are temporally separated. Respectively, computing schemes: every propagating signal leaves an un-removable trace, so quite soon after the computation starts the space will be filled by traces of signals. Having said that, there is experimental evidence (Fig. 13) that when some branches (protoplasmic tubes) of plasmodium become obsolete and cease to function, new pseudopodia can grow over and cross them. This means that the BZ medium and Physarum are both reusable but at different time scales. In both cases though unless nutrients are replenished then the computations will naturally cease at some point. In experimental implementations involving the BZ reaction then reagents are continuously fed into the reactor vessel (however, the catalyst and gel structure are altered very slightly by the wave interactions/ computations meaning that ultimately the computations have a finite lifetime). In Plasmodium experiments nutrients can presumably be added to support growth and overcome excessive nutrient depletion which would affect computation. However, the effect of previous trajectories on current computations is difficult to assess. It is likely that in both the BZ and physarum implementations that initial computations and histories thereof affect all subsequent computations by subtle alterations of wave trajectory and velocity. In chemical systems it is not possible to gauge the effect easily as there is no trace of any previous interaction. As demonstrated in Fig. 14 in Physarum it is more obvious where previous computations affect current computations. The main point is that in both systems the excitability and therefore the computational efficiency are altered by the original wave collisions/ trajectories.

**Speed.** In the BZ system waves propagate with an average speed of 2 mm/min, in plasmodium growing wave-fronts propagate with a speed of 0.1 mm/min. This makes Physarum computer an order slower than BZ computer in structural computation, i.e. propagation in physical space. However, if we compare the oscillation frequencies of micro-volumes in the BZ medium and the electrical activity of the Physarum, we find that the clock rate of Physarum computers is 0.008-0.02 Hz and BZ processors have a clock rate of 0.002 Hz [Adamatzky, De Lacy Costello & Asai, 2005]. We can conclude that both types of computers are therefore equal in speed of computation.

**Memory.** BZ collision-based computers are memory-less. That is to say that although memory effects act to alter BZ collision based computations – they do not possess an addressable memory so this information cannot be easily accessed – only observed in subsequent reactions/collisions. Waves propagate and then medium returns to a state that is macroscopically identical to the initial resting state. In plasmodium growth patterns propagate, and then tree-like structures of protoplasmic tubes are formed. These protoplasmic trees can represent a static memory of Physarum computer.

**Programmability.** BZ and plasmodium are light-sensitive, which give us the means to programme them. Physarum exhibits articulated negative phototaxis, BZ reaction is inhibited by light. Therefore by using masks of illumination one can control dynamics of localizations in these media, change a signal's trajectory or even stop a signal's propagation, amplify the signal, generate trains of signals. Light-sensitive of plasmodium has been already explored in design of robotics controllers [Tsuda et. al., 2007].

**Branching.** Branching is useful when we need to implement signal multiplication. Is the phenomenon observed in the systems discussed? Wave fragments in the BZ reaction do split spontaneously but only in unfavourable conditions such as in light sensitive reactions, where travelling wave fragments may split into two because the light level is just above the sub-excitable threshold (Fig. 14). It may also occur when the daughter fragment of collision result passes back over the refractory tail of parent fragments or another collision/ trajectory site.

The splitting of localizations in BZ medium can be interpreted in terms of a field potential, particularly for the case of light-induced splitting. It is known that at critical uniform light level small fragments of excitation will not travel a large distance without spontaneous splitting. Fragments may get smaller in size in addition to splitting. There are many cases (see Fig. 14) where even though fragments are expanding as they travel across critical light fields then they will also split spontaneously. Exact mechanism is unknown. We can speculate that this is due to heterogeneity in the substrate which affects the chemical environment or light level locally. Usually the distance travelled across the destabilizing field is quite large prior to splitting, maybe the expansion is the cause of lowered stability in already unfavourable conditions.

In Physarum splitting of propagating localization may happen due to heterogeneity of cytoskeleton. Heterogeneous distribution of cytoplasmic streaming promotes heterogeneous arrangement of actin filaments, which enhances the branching in extension of pseudopodia.

**Practicality.** BZ processors function mainly in liquid phase or gel (even experiments on printed BZ medium [Steibock et al., 1995] involved liquid phase of reaction], and very sensitive to disturbance. Also, being left in a closed reactor BZ medium becomes exhausted and ceases functioning.

Plasmodium of *P. polycephalum* can inhabit a wide variety of substrates, including bare glass and aluminium foil because its membrane provides protection from hostile environment thus making internal reaction-diffusion processes less sensitive to disturbances. This makes Physarum computers more suitable for real-world applications.

## 6. Discussions

A field of non-linear media computers [Adamatzky, 2001] – where computation is implemented by patterns propagating in spatially-extended non-linear media – features two main 'competitors'. They are reaction-diffusion chemical computers and Physarum computers. The reaction-diffusion chemical computers are mainly presented by excitable chemical media – Belousov-Zhabotinsky (BZ) processors, and less by precipitating processors, see [Adamatzky, De Lacy Costello & Asai, 2005] for detailed classification. Physarum computers are based on foraging behaviour of and protoplasm transfer minimization inside a vegetative state, or plasmodium, of *Physarum polycephalum* [Nakagaki et al., 2000; Nakagaki, 2001; Nakagaki et al., 2001; Tsuda et al., 2004, 2006; Adamatzky, 2007a].

So far there are several experimental prototypes of reaction-diffusion and Physarum computers that approximate shortest and collision-free paths, Voronoi diagram, skeleton of a planar shape, and control navigation of mobile robots. They perform computation using travelling mobile localizations (local perturbations of medium's characteristics), which collide to each other, change their velocity vectors and/or morphological structures in the result of collision, and thus realize functionally complete sets of logical gates. Prototypes of universal computers (apart of collision-based computers), are the result of collision, and thus realize functionally complete sets of logical gates.

The collision-based computers function in homogeneous, not geometrical constraints imposed, medium. They perform computation using travelling mobile localizations (local perturbations of medium's characteristics), which collide to each other, change their velocity vectors and/or morphological structures in the result of collision, and thus realize functionally complete sets of logical gates.

Both BZ and Physarum computers exhibit travelling localizations: wave-fragments in BZ medium and pseudopodia in plasmodium of Physarum. We have provided a detailed comparison of the mobile localization in these two types of non-linear medium computers. We have demonstrated that despite significant physico-chemical difference – e.g. BZ media are homogeneous and Physarum have optional 'hardware' like cytoskeleton and membrane – spatio-temporal dynamics of localization is similar. At some level of abstraction tips of pseudopodia in Physarum and travelling excitations in BZ medium show the same range of characteristics. Advantages and disadvantages of BZ and Physarum computers — only from collision-based computing point of view -- are summarized in Table 1. BZ computers are preferable in terms of richness of realizable operations (measured in number of different outcomes of collisions between mobile localizations) and architectural capacity (evaluated on a number of different dynamical architectures embeddable into the medium). Physarum computers are winners in terms of memory implementation and practicality.

What would be the best substrate for implementation of future collision-based computers – plasmodium of *Physarum polycephalum* or Belousov–Zhabotinsky medium? There is no absolute answer. Advantageous feature of the substrates are complementary, and a choice of exact implementation of one or another computing scheme should always depend on a problem domain.

## Acknowledgements

We are in debt to Dr. Soichiro Tsuda for commenting on various aspects of plasmodium behaviour, particularly on interactions of pseudopodia. A.A. is also expressing his sincere gratitude to Dr. Tsuda for initiating him in the field of Physarum computing.

---

**FIGURES**

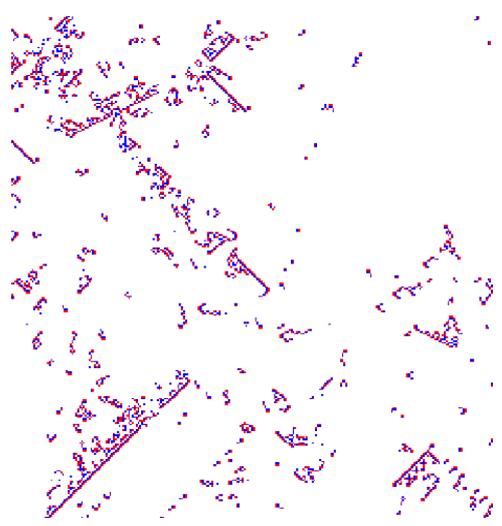

**Figure 1**. Snapshots of excitation dynamics in orthogonal excitable lattice where every resting site excites if there are exactly two excited neighbours. We can see many localized excitations travelling on the lattice, and also generators of excitations.

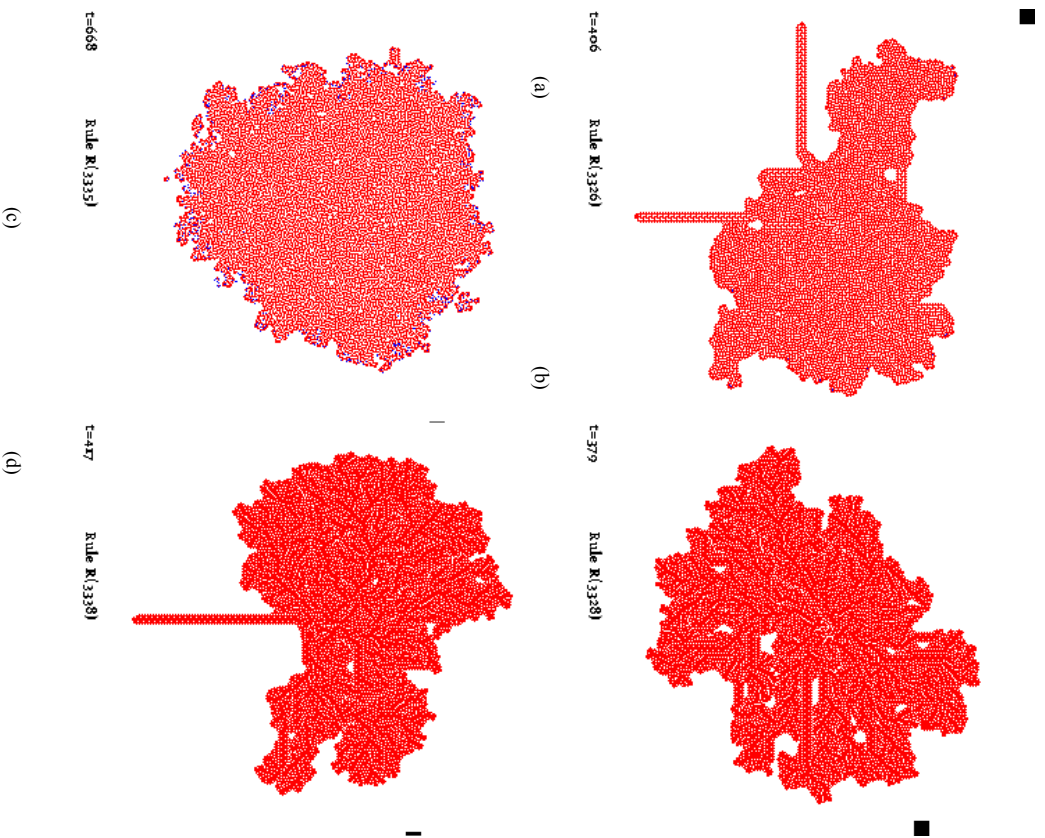

t=406    Rule R(3326)          t=379    Rule R(3328)

(a)                          (b)

t=668    Rule R(3335)          t=447    Rule R(3338)

(c)                          (d)

**Figure 2** Snapshots of cellular automaton with retained excitation rules. For each automaton development in a circle radius 15 cells, assigned states at random. (a) $R(3,3,2,6)$, (b) $R(3,3,2,8)$, (c) $R(3,3,3,5)$, (d) $R(3,3,3,8)$.

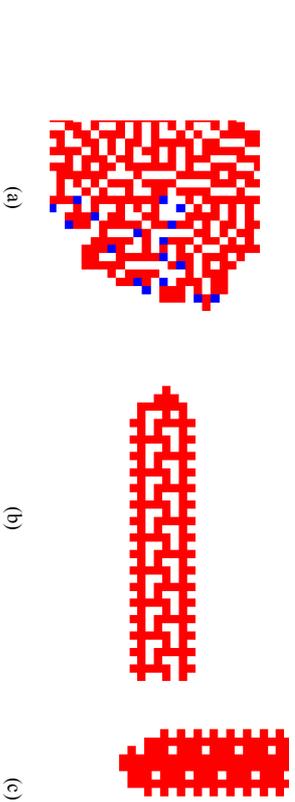

(a)

(b)

(c)

**Figure 3.** Growth-points in cellular automaton with retained excitation rules. (a) $R(3,3,3,5)$, (b) $R(3,3,2,6)$, (c) $R(3,3,3,8)$

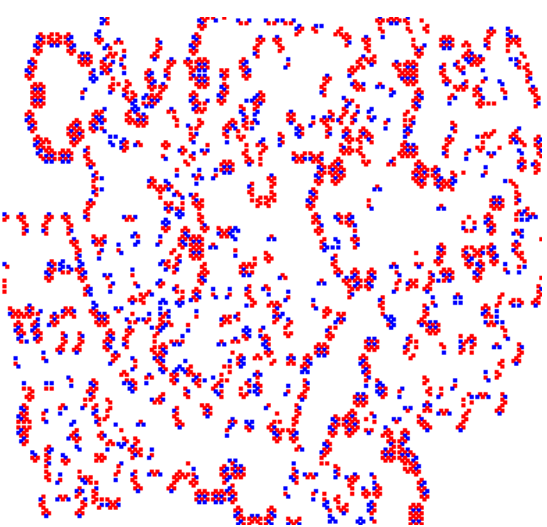

**Figure 4.** Dynamics of localizations in hexagonal reaction-diffusion cellular automaton, imitating the following set of quasi-chemical reactions (see [Adamatzky et al., 2007] for details of converting cell-state transition rule into a set of reactions):

$B \to A$
$3B \to A$
$A + B \to B$
$A + 2B \to A$
$3A + 3B \to A$

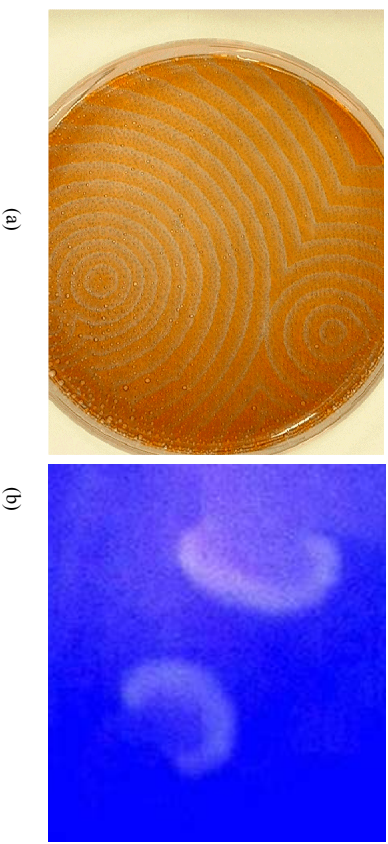

**Figure 5.** Two types of wave dynamics in experimental Belousov-Zhabotinsky system: (a) classical target waves in excitable medium, (b) travelling localizations, or wave-fragments, in sub-excitable medium.

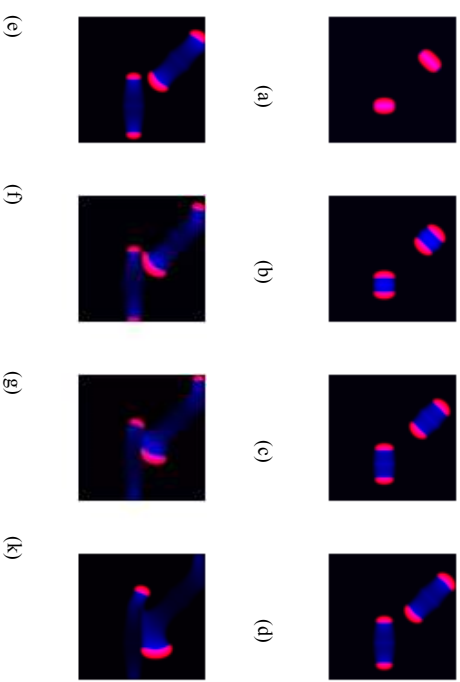

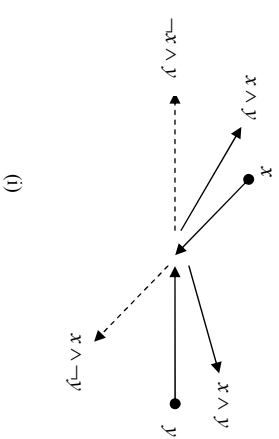

**Figure 6.** Examples of a logical gate implemented in collision of travelling localizations (wave-fronts) in numerical model of Belousov-Zhabotinsky system: (a)—(h) snapshots of wave-fragment dynamics; (i) schematic representation of a gate. The two-variable Oregonator equations were integrated using Euler method with five node Laplacian operator, time step $\Delta t = 10^{-3}$ and grid point spacing $\Delta x = 0.15$ with the following parameters: $\varepsilon = 0.03$, $f = 1.4$, $q = 0.002$, $\Phi = 0.04$. The dynamics of collision-based gate in (a)—(h) represent the case when both input variables have values Truth. These values were represented by local disturbances of initial concentrations of species.

(a)

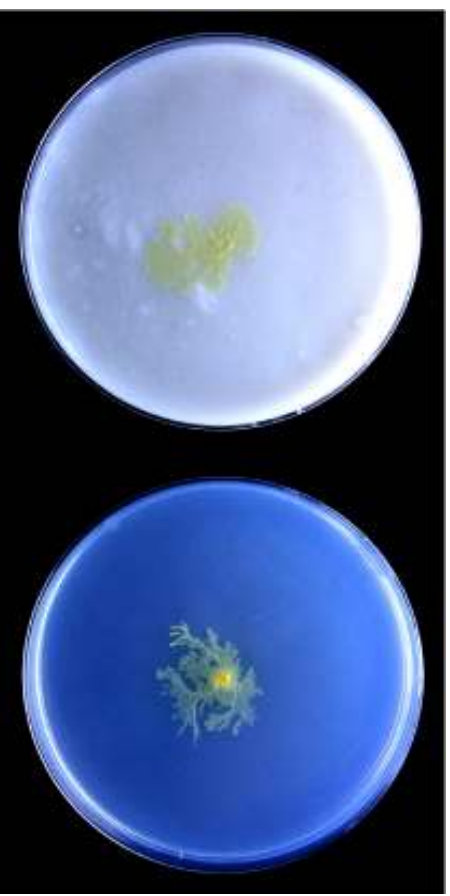
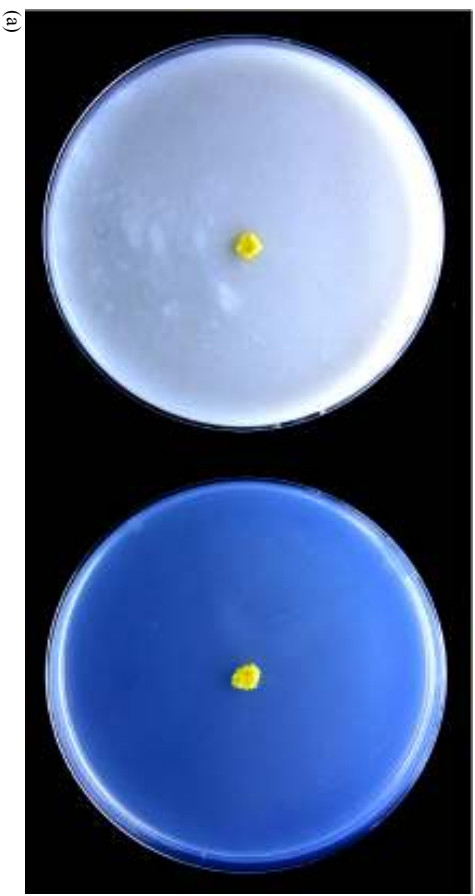

(b)

Figure 7 (a) 20 mg of plasmodia were placed on 50 mg/ml crushed oat flakes containing 1.5 % agar gel (left, nutrient-rich condition) or 100 mM potassium chloride containing 1.5 % agar gel (right, no-nutrient and with repellent potassium). (b) Configurations of plasmodium developed 5.5 hours later. In the nutrient rich favourable condition the plasmodium formed circular pseudopodia, on the other hand in unfavourable condition the plasmodium developed branching structure.

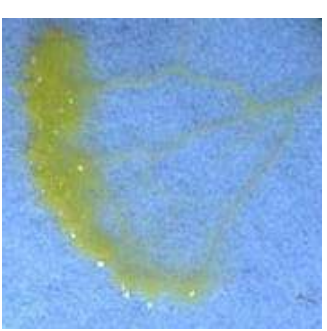

Figure 8. Typical travelling localization in plasmodium of *P. polycephalum* on nutrient poor substrate. Plasmodium was placed on wet filter paper, oat flakes very sparsely scattered on the filter paper represented sources of nutrients.

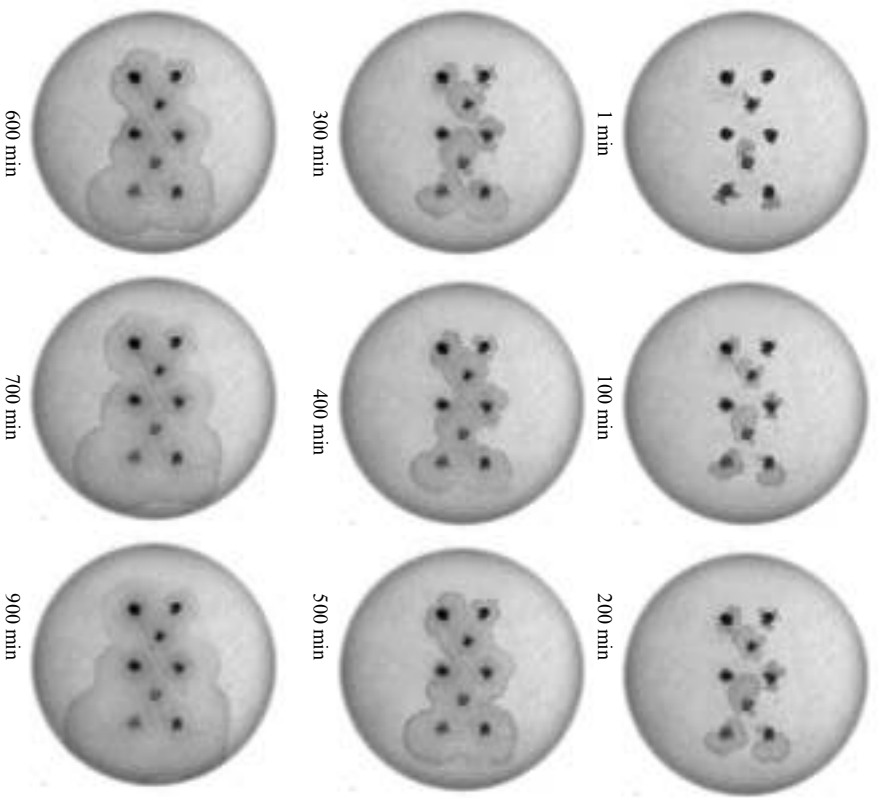

**Figure 9.** Propagation of plasmodium of *P. polycephalum* in nutrient-rich, 50 mg/ml crushed oat flake containing 1.5 % agar gel substrate.

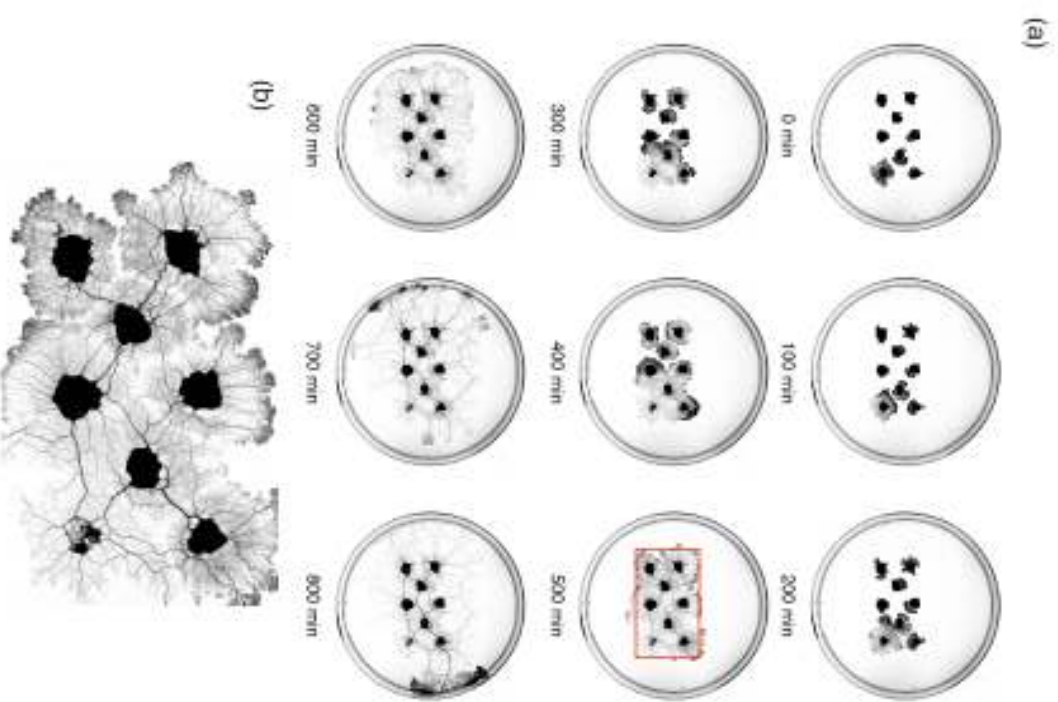

**Figure 10** (a) Propagation of plasmodium of *P. polycephalum* in a non-nutrient bare agar gel substrate. (b) Magnified image of the area within the red rectangle in (a). Though the pieces are connected by bypassing tubes, pseudopodia avoid each other, as the gap between the pieces indicates.

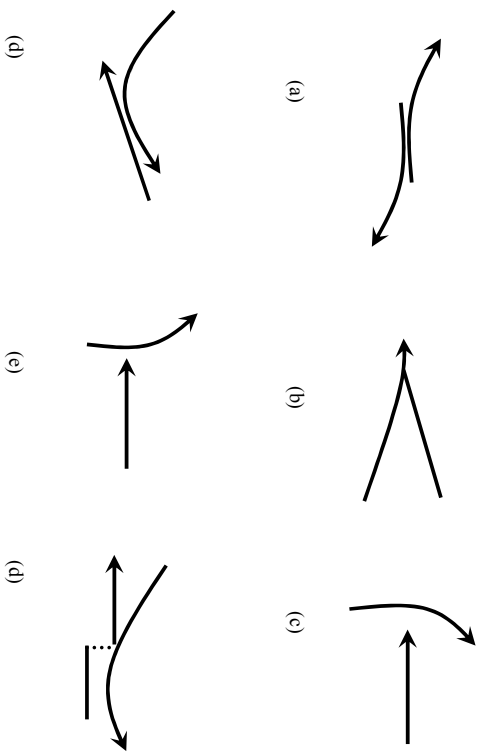

**Figure 11.** Adamatzky-De Lacy Costello classification of binary collisions between mobile localizations in BZ medium: (a) reflection, (b) fusion, (c) attraction, (d) sliding, (e) repulsion, (d) shifting-sliding [Adamatzky & De Lacy Costello, 2007].

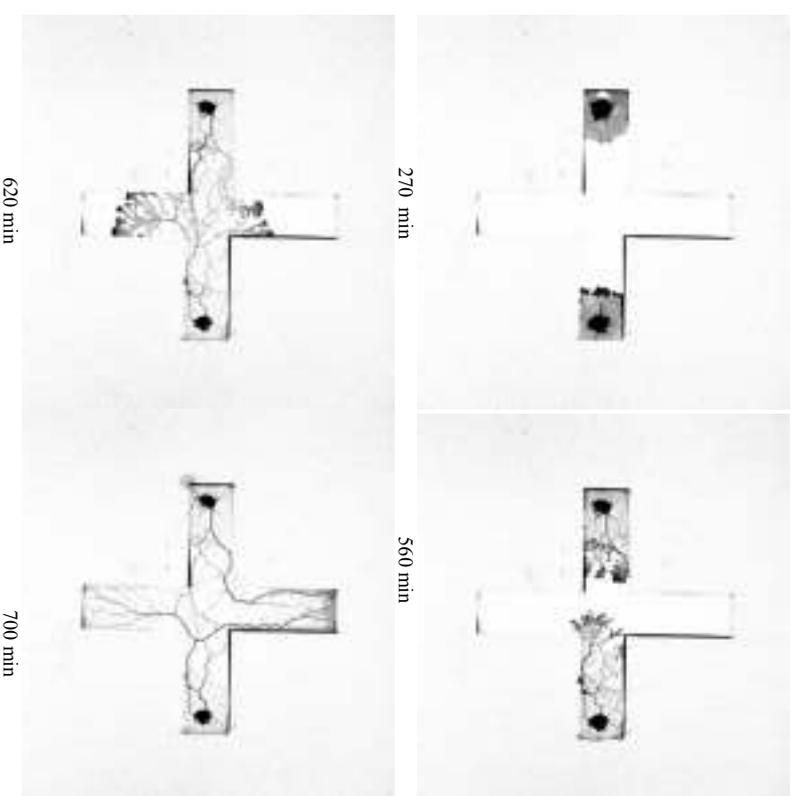

Figure 12. Interaction of pseudopodia of plasmodium: 1 mg of plasmodia were placed at opposite sides of the horizontal channel, 1.5% agar gel was a non-nutrient substrate.

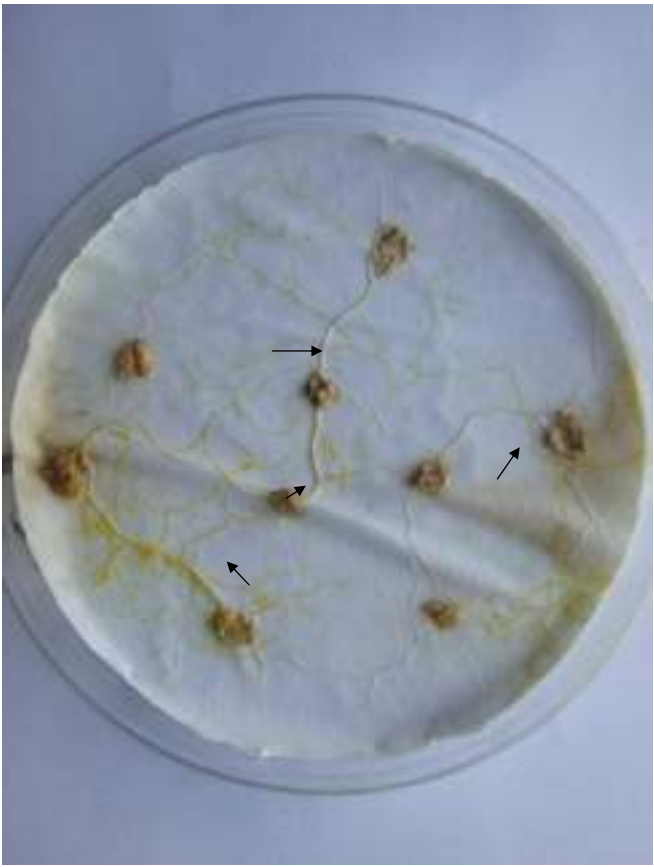

**Figure 13.** Demonstration of possible space reusing by plasmodium. Places where living pseudopodia cross over dead ones are shown by arrows. Photography of plasmodium of *Physarum polycephalum* cultivated on wet filter paper. Environment is nutrient poor. Oat flakes are clearly visible on the photography.

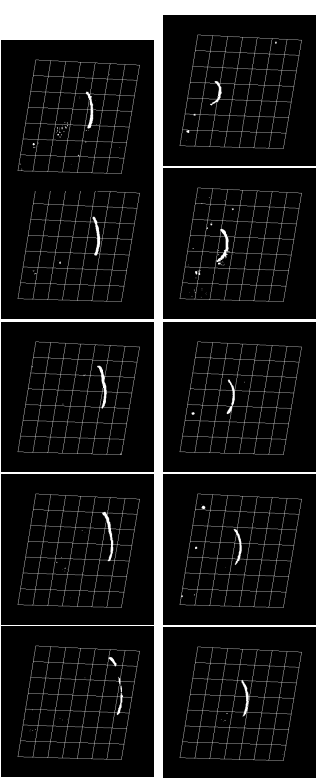

**Figure 14.** A series of snapshots of experimental BZ medium, demonstrating splitting/branching of wave-fragments. Snapshots are ordered chronologically from the left to the right, and from the top to the bottom.

| Characteristic | BZ | Physarum |
|---|---|---|
| Richness of outcomes of localization collisions | + | - |
| Architectural capacity | + | - |
| Memory | - | + |
| Speed of operations | + | + |
| Programmability | + | + |
| Practicality | - | + |

**Table 1.** Comparative analysis of BZ and Physarum collision-based computers. Symbol "-" indicates inferior characteristics.